\documentclass[12pt]{article}
\usepackage{epsf,epsfig,amssymb,amsmath,graphicx,float}
\usepackage{color}

\begin{document}
\renewcommand{\thefootnote}{\fnsymbol{footnote}}

\begin{titlepage}

\begin{center}

\vspace{1cm}

{\Large {\bf Asymmetric Dark Matter in the Shear--dominated
    Universe }}

\vspace{1cm}

{\bf Hoernisa Iminniyaz}\footnote{wrns@xju.edu.cn}
\vskip 0.15in
{\it
%$^a$
{School of Physics Science and Technology, Xinjiang University, \\
Urumqi 830046, China} \\

}

\abstract{ We explore the relic abundance of asymmetric Dark Matter in
  shear--dominated universe in which it is assumed the universe is
  expanded anisotropically. The modified expansion rate leaves its imprint on
  the relic density of asymmetric Dark Matter particles if the asymmetric Dark
  Matter particles are decoupled in shear dominated era. We found the relic
  abundances for particle and anti--particle are increased. The particle and
  anti--particle abundances are almost in the same amount for the larger value
  of the shear factor $x_e$ which makes the indirect detection possible
  for asymmetric Dark Matter. We use the present day Dark Matter density from
  the observation to find the constraints on the parameter space
  in this model. }
\end{center}
\end{titlepage}
\setcounter{footnote}{0}

\section{Introduction}

In addition to the Wilkinson Microwave Anisotropy Probe (WMAP) data
\cite{wmap}, the Planck mission provided the value of the Dark Matter relic
density with high precision recently \cite{Ade:2015xua}. Planck 2015 data
gives the present cold Dark Matter relic density as
\begin{eqnarray} \label{data}
  \Omega_{\rm DM} h^2 = 0.1199 \pm 0.0022\, ,
\end{eqnarray}
where $h = 0.673 \pm 0.098$ is the present Hubble
expansion rate in units of 100 km s$^{-1}$ Mpc$^{-1}$ \cite{Ade:2015xua}.

Although there are strong evidences for the
existence of Dark Matter, the nature of the Dark Matter is still not known. The
usual assumption is that the neutral, long--lived or stable Weakly
Interacting Massive Particles (WIMPs) are the best motivated candidates for
Dark Matter. Neutralino is one example which is appeared in supersymmetry
and it is Majorana particle for which
its particle and anti--particle are the same. However, there are other
possibilities that the Dark Matter can be asymmetric which means the particles
and anti–particles are distinct from each other if the particles are fermionic
\cite{adm-models,frandsen}.

The relic density of asymmetric Dark Matter in the standard cosmological
scenarios and non-standard cosmological scenarios like quintessence,
scalar-tensor and brane world cosmological scenarios are discussed in
\cite{GSV,Iminniyaz:2011yp,Gelmini:2013awa,Iminniyaz:2013cla,Meehan:2014zsa,Wang:2015gua}. In
nonstandard cosmological scenarios, the Hubble expansion rate is changed
comparing to the standard cosmological scenario. If the asymmetric Dark Matter
particles decay during the era in which the expansion rate is changed, both
the Dark Matter particle and anti--particle abundance are affected by the
modification.

The particle and anti--particle abundances are determined by solving the
Boltzmann equations which are the evolution equations of the particles and
anti--particles in the expanding universe. Several nonstandard cosmological
models \cite{Gelmini:2013awa,Iminniyaz:2013cla,Meehan:2014zsa,Wang:2015gua}
discussed the effect of the modified
expansion rate on the relic density of asymmetric Dark Matter. The
characteristic of the nonstandard cosmological models which are discussed in
\cite{Gelmini:2013awa,Iminniyaz:2013cla,Meehan:2014zsa,Wang:2015gua} is
the increase of the particle and anti--particle abundance due to
the increased Hubble expansion rate. In nonstandard cosmological scenarios,
for appropriate annihilation cross section, the deviation of the
abundance between the particle and anti--particle are not large. For
asymmetric Dark Matter, in the beginning we assume there are more particles
than the
anti--particles; in the end the anti--particles are completely annihilated away
with the particles and there are no anti--particles left. This makes the
indirect detection is impossible for asymmetric Dark Matter in the standard
cosmological scenarios. However, it is changed for non-standard cosmological
scenarios. The increased annihilation rate for particle and anti--particle
provides us the possibility that the
that the asymmetric Dark Matter can be detected indirectly.

One of the interesting nonstandard cosmological model is the Bianchi type I
model in which it is
assumed the expansion of the universe is not isotropic. In this model the
effects of the anisotropy on the expansion rate of the universe is quantified
by an anisotropy--energy density which decreases as $R^{-6}$
\cite{Bianchi1,Bianchi2,Bianchi}. In
\cite{Kamionkowski:1990ni,Barrow:1982ei}, the authors
investigated the relic abundance
of Dark Matter in Bianchi I cosmological model. In
this paper we extend the discussion of
\cite{Kamionkowski:1990ni,Barrow:1982ei} to the asymmetric
Dark Matter case. We investigated how the abundance of asymmetric Dark Matter
particles can be affected if the
asymmetric Dark Matter particles are decayed in shear--dominated
universe. We discuss in detail the relic density of asymmetric Dark Matter in
shear--dominated universe and then use the present Dark Matter density from the
observation to find the constraints on the
parameter space in shear-dominated model.

The arrangement of the paper is following. In section 2, we review the
shear--dominated universe. The asymmetric Dark Matter
relic density is discussed both in numeric and analytic way in
shear--dominated universe in section 3. In section 4, the constraints
on the parameter space of shear--dominated universe are obtained using the
observed Dark Matter relic density. The conclusion and summary are in the
final section.

\section{Review of the shear--dominated universe}

In this section, we review the shear--dominated universe. In the standard
cosmological scenario, it is assumed that the universe is
isotropic and homogeneous. However before Big Bang Nucleosynthesis (BBN),
there is no evidence which shows that the universe should be homogeneous or
isotropic. One of the nonstandard cosmological scenarios is Bianchi type I
model which is homogeneous but anisotropic cosmological model
\cite{Bianchi1,Bianchi2,Bianchi}. The expansion rate in this scenario is
\begin{equation}
      H^2 = \frac{8\pi G}{3 } (\rho_r + \rho_s),
\end{equation}
where $G = 1/(8\pi M^2_{\rm Pl})$ with $M_{\rm Pl} = 2.4 \times 10^{18}$ GeV,
$\rho_r = \pi^2 g_* \,T^4/30 $ is the radiation energy density,
here $g_*$ is the effective number of the relativistic degrees of freedom.
$\rho_s$ is the shear energy density which is defined to be
\begin{equation}
      \rho_s \equiv \frac{1}{48 \pi G} [(H_1 - H_2)^2 + (H_1 - H_3)^2 +
        (H_2- H_3)^2]\, ,
\end{equation}
where $H_i \equiv \dot{R_i}/R_i $ are the expansion rates for the three
principal axes with $R_i$ being the scale factors of the three principal axes
of the universe. It is derived that shear energy density is proportional to
$\rho_s \propto \bar{R}^{-6} $ \cite{Kamionkowski:1990ni}. Here $\bar{R}$
is the mean--scale
factor as $\bar{R} = V^{1/3}$ with $V = R_1 R_2 R_3$. We need explicit
expression for the expansion rate to calculate the asymmetric Dark Matter
relic density in this model. Using the conservation of the entropy per
co--moving volume $g_* \bar{R}^3 T^3 = {\rm const} $, the shear--energy
density is expressed as $\rho_s \propto g^2_* T^6/{\rm const}$. It is
defined at temperature $T_e$, $\rho_r = \rho_s$. The universe is shear
dominated when
$T \gg T_e$, in which $H \propto \bar{R}^{-3} $ and $\bar{R} \propto t^{1/3}$;
when the temperature falls well below $T_e$ as $T \ll T_e$, the universe is
radiation dominated, here the expansion rate $ H \propto \bar{R}^{-2}$ and
$\bar{R} \propto t^{1/2}$. In terms of the radiation--energy density, the
shear--energy density can be written as
\begin{equation}
      \rho_s = \rho_r \left[ \frac{g_* T^2}{g^e_* T_e^2} \right],
\end{equation}
where $g_*^e $ is the value of $g_*$ at $T_e$. The shear--energy density must
be sufficiently small in order
not to conflict with the successful prediction of
BBN. BBN imposed the bounds on $T_e \geq 2.5$ MeV.
Then the total energy density in shear dominated universe is
\begin{equation}
 \rho = \rho_r + \rho_s =\frac{\pi^2}{30} g_* T^4
      \left[ 1 + \frac{g_* T^2}{ g^e_* T^2_e}  \right]
\end{equation}
Finally the modified expansion rate is
\begin{equation} \label{eq:Hubble}
      H =  \frac{\pi m_{\chi}^2}{ M_{\rm Pl}\, x^2}
          \sqrt{\frac{g_*}{90}\left(1 + \frac{x_e^2}{x^2}\right)}\,,
\end{equation}
where $x = m_{\chi}/T$ with $m_{\chi}$ being the mass of Dark Matter particles
and the shear factor $x_e$ is defined to be
\begin{equation}
      x_e \equiv \frac{m_{\chi}}{T_e} \left[ \frac{g_*}{g^e_*} \right]^{1/2}\, .
\end{equation}

\section{Freeze--out of asymmetric Dark Matter in shear--dominated universe}
The modified Hubble expansion rate has affects on the relic density of
asymmetric Dark Matter in shear--dominated universe.
In this section we investigate to what extent the relic density of asymmetric
Dark Matter is affected if the asymmetric Dark Matter particles freeze--out in
shear--dominated era. Dark Matter relic density is obtained by solving the
following Boltzmann equations for particle and anti--particle which are
written with the modified expansion rate as
\begin{eqnarray} \label{eq:boltzmann_n}
\frac{{\rm d}n_{\chi}}{{\rm d}t} + 3 H n_{\chi} &=&
        - \langle \sigma_{\chi\bar\chi} v\rangle
  (n_{\chi} n_{\bar\chi} - n_{\chi,{\rm eq}} n_{\bar\chi,{\rm eq}})\,;
  \nonumber \\
\frac{{\rm d}n_{\bar\chi}}{{\rm d}t} + 3 H n_{\bar\chi} &=&
   - \langle \sigma_{\chi\bar\chi} v\rangle (n_{\chi} n_{\bar\chi} - n_{\chi,{\rm
       eq}} n_{\bar\chi,{\rm eq}})\,,
\end{eqnarray}
where $\chi$ is the Dark Matter particle which is {\em not} self--conjugate,
i.e. the anti--particle $\bar\chi \neq \chi$. In our work, we assumed that
only $\chi \bar \chi$ pairs can annihilate into
Standard Model (SM) particles. Therefore $\langle\sigma_{\chi\bar\chi} v\rangle$
is the thermal average of the cross section of the annihilating particles
$\chi$ and anti--particles $\bar\chi$ multiplied with the velocity of the
annihilating particles. Here
$n_{\chi}$ and $n_{\bar\chi}$ are the number densities of particle and
anti--particle and their equilibrium values are
$n_{\chi,{\rm eq}} = g_\chi ~{\left( m_\chi T/2 \pi \right)}^{3/2}
{\rm e}^{(-m_\chi + \mu_\chi)/T}$ and
$n_{\bar{\chi},{\rm eq}}=g_\chi ~{\left( m_\chi T/2 \pi
\right)}^{3/2} {\rm e}^{(-m_\chi - \mu_{\bar\chi})/T}$.
Here we assume that the
asymmetric Dark Matter particles were non--relativistic at decoupling.
$\mu_\chi$, $\mu_{\bar\chi}$ are the chemical potential of the
particle and anti--particle, $\mu_{\bar\chi} = -\mu_\chi$ in equilibrium.

We follow the same method as in
\cite{Iminniyaz:2011yp} and
obtain the number densities for particle and anti--particle in shear--dominated
universe. We assume the asymmetric Dark
Matter particles $\chi$ and $\bar{\chi}$ were in thermal equilibrium when the
temperature is high in the early universe. When $T \leq m_{\chi}$, the
equilibrium values of the number densities $n_{\chi,{\rm eq}}$,
$n_{\bar\chi,{\rm eq}}$ decrease exponentially for $m_\chi >
|\mu_\chi|$. Later the interaction rates for particle
$\Gamma =  n_{\chi} \langle \sigma_{\chi\bar\chi} v \rangle$ and anti--particle
$\bar{\Gamma} = n_{\bar\chi} \langle \sigma_{\chi\bar\chi} v \rangle $ drop below the
expansion rate $H$. This process leads to the decoupling of the particles and
anti--particles from the thermal bath and the co--moving number densities are
almost fixed from that inverse--scaled freeze--out temperature $x_F$.

The Boltzmann equations (\ref{eq:boltzmann_n}) can be rewritten
in terms of the dimensionless quantities $Y_\chi =n_\chi/s$,
$Y_{\bar\chi} = n_{\bar\chi}/s$ and $x = m_\chi/T$, where
$ s= (2 \pi^2/45) g_* T^3 $ is the entropy density.
Inserting Eq.(\ref{eq:Hubble}) to the
Boltzmann equations (\ref{eq:boltzmann_n}), then
\begin{equation} \label{eq:boltzmann_Y}
\frac{d Y_{\chi}}{dx} =
- \frac{\lambda \langle \sigma_{\chi\bar\chi} v \rangle}{x\,\sqrt{x^2 + x^2_e}}~
     (Y_{\chi}~ Y_{\bar\chi} - Y_{\chi, {\rm eq}}~Y_{\bar\chi, {\rm eq}}   )\,;
\end{equation}
\begin{equation} \label{eq:boltzmann_Ybar}
\frac{d Y_{\bar{\chi}}}{dx}
= - \frac{\lambda \langle \sigma_{\chi\bar\chi} v \rangle}{x\,\sqrt{x^2 + x^2_e}}~
 (Y_{\chi}~Y_{\bar\chi} - Y_{\chi, {\rm eq}}~Y_{\bar\chi, {\rm eq}} )\,,
\end{equation}
where $\lambda = 1.32\,m_{\chi} M_{\rm Pl}\, \sqrt{g_*}$.
Combining these two equations (\ref{eq:boltzmann_Y}), (\ref{eq:boltzmann_Ybar}),
we obtain
\begin{equation}  \label{eq:epsilon}
 Y_{\chi} - Y_{\bar\chi} = \varepsilon\,,
\end{equation}
here $\varepsilon$ is constant.
Then the Boltzmann equations (\ref{eq:boltzmann_Y}) and
(\ref{eq:boltzmann_Ybar}) become
\begin{equation} \label{eq:Yc}
\frac{d Y_{\chi}}{dx} =
  -\frac{\lambda \langle \sigma_{\chi\bar\chi} v \rangle}{x\,\sqrt{x^2 + x^2_e}}~
                 (Y_{\chi}^2 - \varepsilon Y_{\chi} - P     )\, ;
\end{equation}
\begin{equation} \label{eq:Ycbar}
\frac{d Y_{\bar{\chi}}}{dx} =
  - \frac{\lambda \langle \sigma_{\chi\bar\chi} v \rangle}{x\,\sqrt{x^2 + x^2_e}}
 (Y_{\bar\chi}^2 + \varepsilon Y_{\bar\chi}  - P)\,,
\end{equation}
where $P$ is the product of $ Y_{\chi, {\rm eq}}$ and
$ Y_{\bar\chi,{\rm eq}}$,
$P = Y_{\chi, {\rm eq}} Y_{\bar\chi,{\rm eq}} = (0.145g_{\chi}/g_*)^2\, x^3 e^{-2x}$.
To solve these Boltzmann equations (\ref{eq:Yc}) and
(\ref{eq:Ycbar}), we use the annihilation cross section which is expanded
in the relative velocity $v$ of the annihilating Dark Matter particles,
the thermal average is
\begin{equation}
\langle \sigma_{\chi\bar\chi} v \rangle =a + 6\,b x^{-1} + {\cal O}(x^{-2}),
\end{equation}
where $a$ is the $s$ wave contribution to $\sigma v$ when $v \rightarrow 0$
and $b$ is the $p$--wave contribution when $s$--wave annihilation is suppressed.
\begin{figure}[h!]
  \begin{center}
    \hspace*{-0.5cm} \includegraphics*[width=8cm]{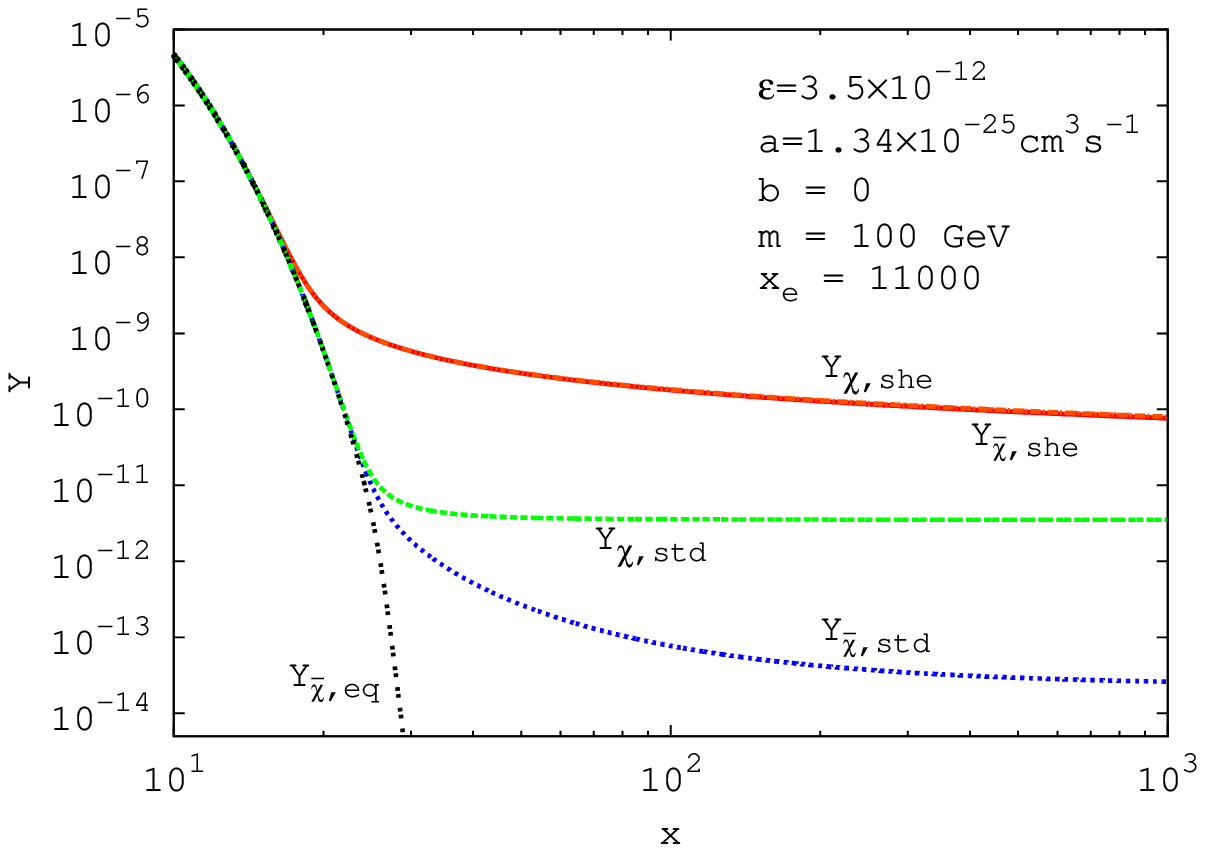}
    \put(-115,-12){(a)}
    \hspace*{-0.5cm} \includegraphics*[width=8cm]{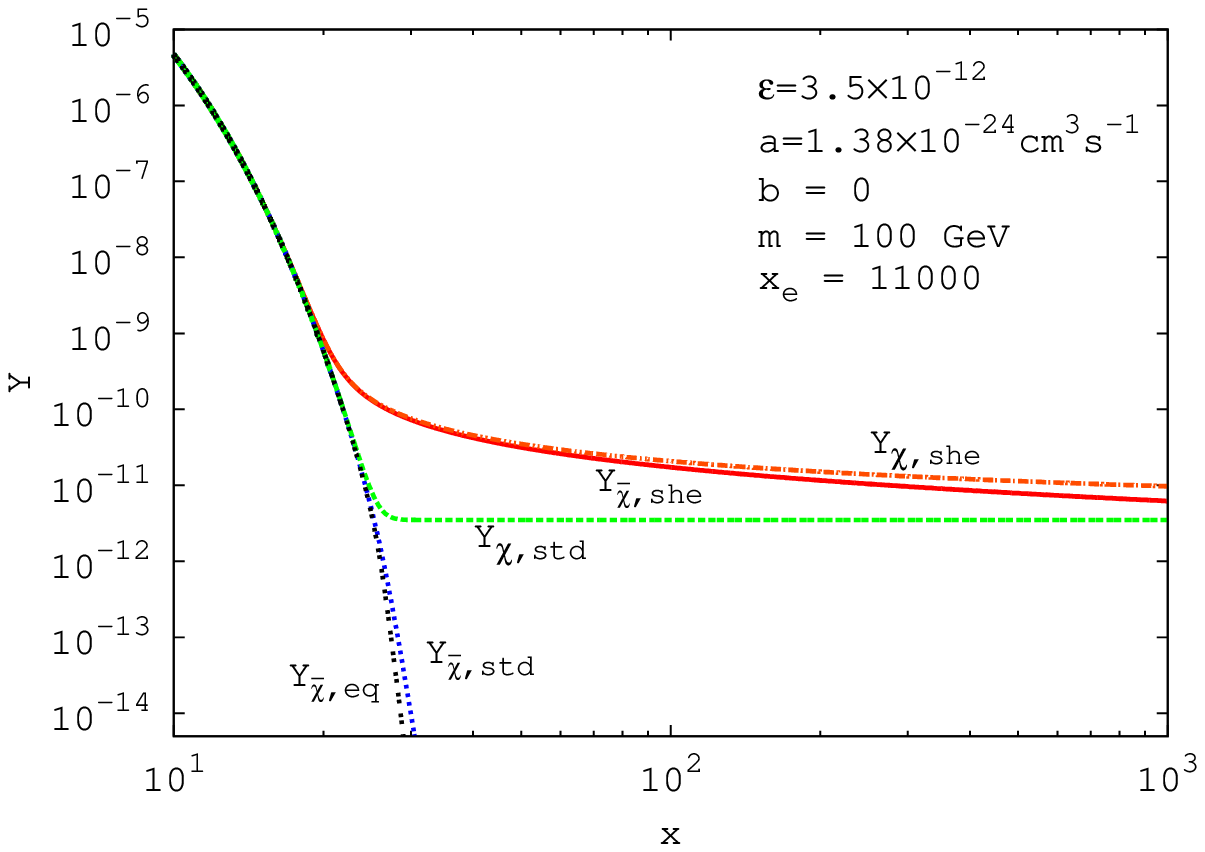}
    \put(-115,-12){(b)}
    \vspace{0.2cm}
    \hspace*{-0.5cm} \includegraphics*[width=8cm]{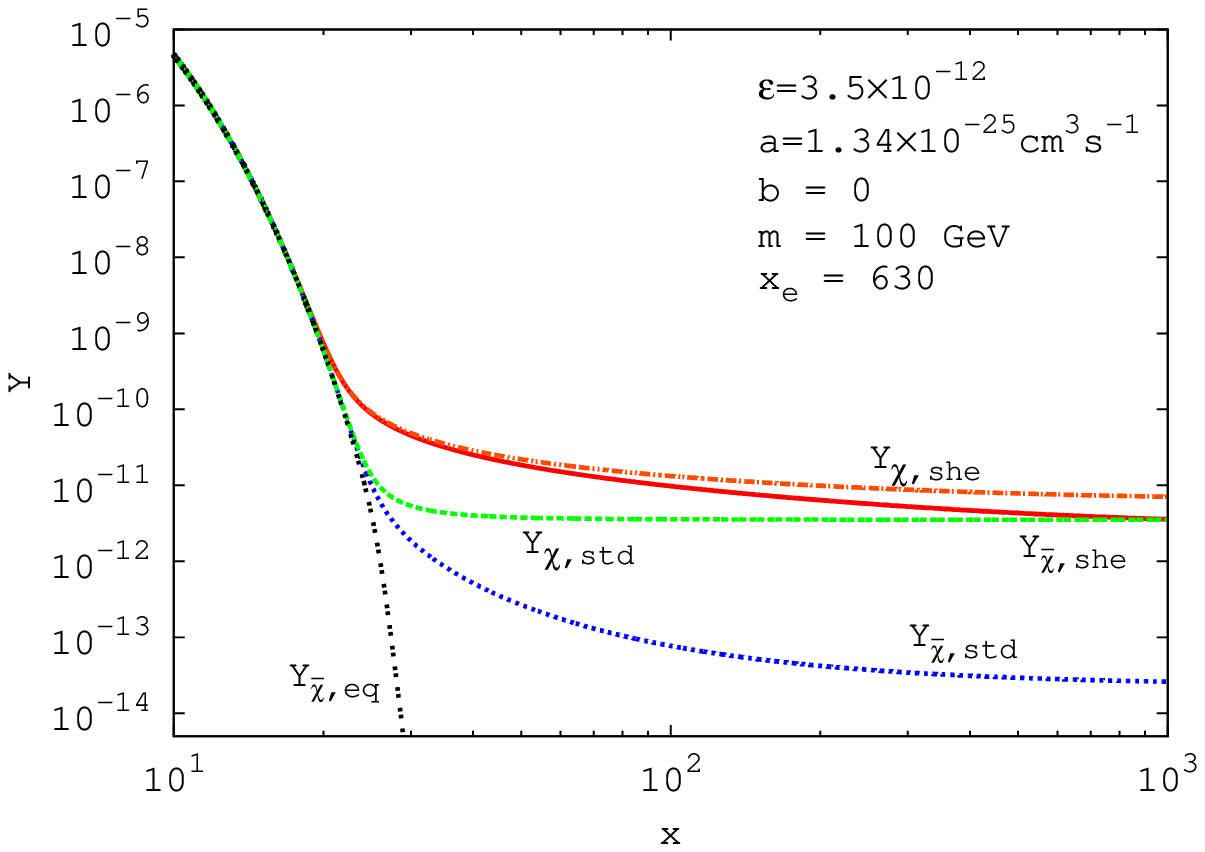}
    \put(-115,-12){(c)}
    \hspace*{-0.5cm} \includegraphics*[width=8cm]{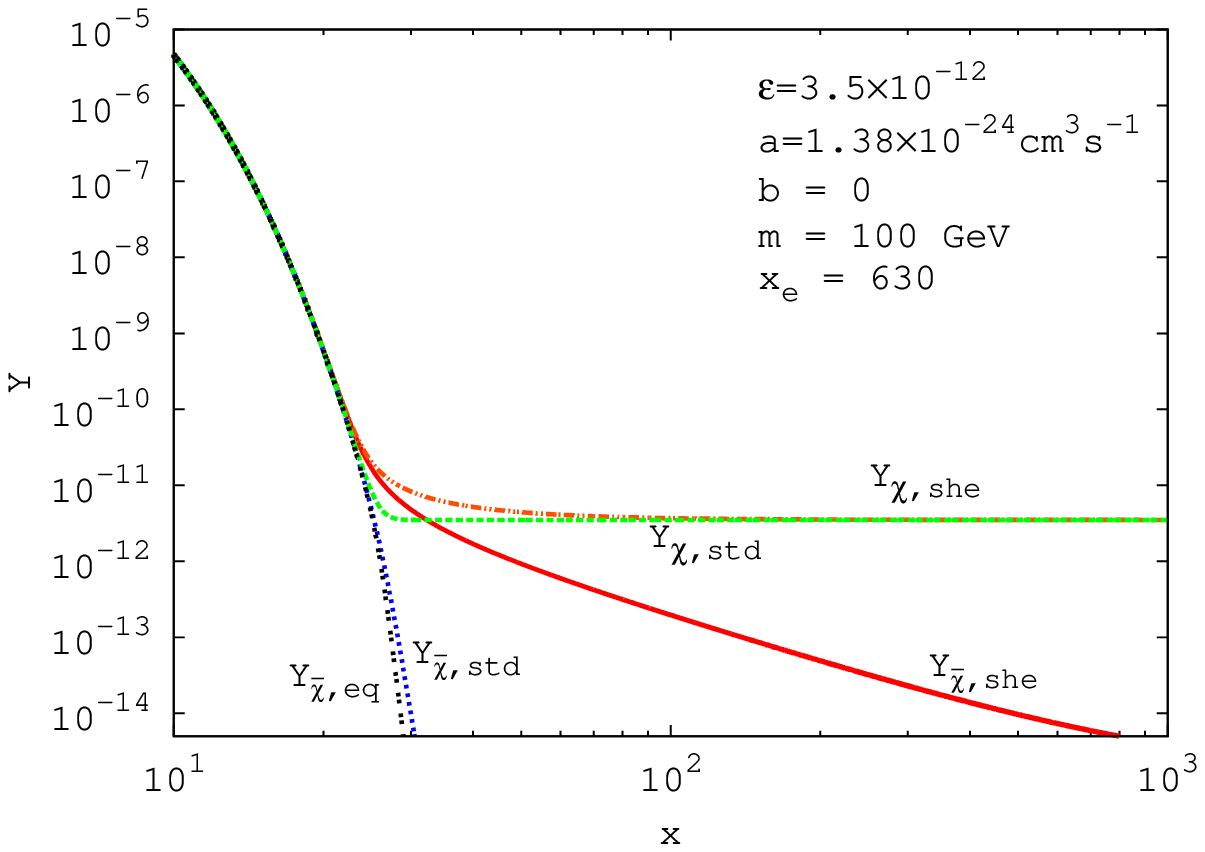}
    \put(-115,-12){(d)}
     \caption{\label{fig:a} \footnotesize The relic abundances $Y_{\chi}$ and
     $Y_{\bar\chi}$ for particle and anti--particle in the standard
       and shear--dominated universe as a
     function of the inverse--scaled temperature $x$ for
     $ \varepsilon = 3.5 \times 10^{-12} $, $m_\chi = 100$ GeV, $g_{\chi} = 2$,
     $g_* = 90$, $b = 0$, $ a = 1.34 \times 10^{-25} $cm$^3$ s$^{-1}$ for
     (a) and (c), $ a = 1.38 \times 10^{-24} $cm$^3$ s$^{-1}$ for
     (b) and (d).
     $x_e = 11000$ for panels (a), (b) and $x_e = 630$ for panels
     (c) and (d). }
  \end{center}
\end{figure}
Using the numerical solution of Eqs.(\ref{eq:Yc}), (\ref{eq:Ycbar}), we plot
the evolution of $Y_{\chi, \bar\chi}$ as a
function of the inverse--scaled temperature $x$ which is shown in
Fig.\ref{fig:a}.
The dot--dashed (red) line is the abundance $Y_{\chi,{\rm she}}$ for particle
and thick (red) line is the abundance $Y_{\bar\chi,{\rm she}}$ for anti--particle in
shear--dominated universe; the dashed
(green) line is the abundance $Y_{\chi,{\rm std}}$ for particle and dotted
(blue) line is the abundance $Y_{\bar\chi,{\rm std}}$ for anti--particle in
the standard
scenario. The double dotted (black) line is the equilibrium value of the
anti--particle abundance $Y_{\bar\chi,{\rm eq}}$. In this figure we can see
how the shear increases the abundances of particle and anti--particle in the
same time. For the same cross section and same asymmetry as in panels (a) and
(c), the effect is more sizable for large $x_e$. Both the particle and
anti--particle abundances are increased in
the shear--dominated universe. We see that the particle and anti--particle
decays earlier in shear--dominated
model than the standard model, this leads to the increase of the abundances.
The increase depends on the value of $x_e$ for
the same cross section. For example, the
increase is larger for $x_e = 11000$ in panel (a)
comparing to the case in panel (c).
For the same shear factor $x_e$, the increase is small for the larger
annihilation cross section in shear--dominated universe. It is shown in
panels (a) and (b).
For $x_e = 630$, the particle abundance in shear--dominated universe is almost
same with the abundance in the standard cosmological scenario when the cross
section is as large as $ a = 1.38 \times 10^{-24} $cm$^3$ s$^{-1}$. We see
that the anti--particle abundance is still suppressed for large annihilation
cross section as in panel (d).

We can have the analytic solution of the relic density for asymmetric Dark
Matter in shear--dominated model repeating the same method as in
\cite{Iminniyaz:2011yp}. We only need to solve Eq.(\ref{eq:Ycbar}) for
$\bar{\chi}$ density. Later we can easily find $Y_{\chi}$ using the
relation $Y_{\chi} - Y_{\bar\chi} = \varepsilon$. In terms of
$\Delta_{\bar\chi} = Y_{\bar\chi} - Y_{\bar\chi,{\rm eq}}$, the Boltzmann
equation (\ref{eq:Ycbar}) is rewritten as
\begin{equation} \label{eq:delta}
\frac{d \Delta_{\bar\chi}}{dx} = - \frac{d Y_{\bar\chi,{\rm eq}}}{dx} -
\frac{\lambda \langle \sigma_{\chi\bar\chi} v \rangle}{x\sqrt{x^2 + x^2_e}}~
\left[\Delta_{\bar\chi}(\Delta_{\bar\chi} + 2 Y_{\bar\chi,{\rm eq}})
      + \varepsilon \Delta_{\bar\chi}   \right]\, .
\end{equation}
We get the solutions of this equation for two extreme cases: the solution
for high temperature is
\begin{equation} \label{bardelta_solu}
      \Delta_{\bar\chi} \simeq \frac{2 P x^2 \sqrt{x^2 + x^2_e}}
   {\lambda \langle \sigma_{\chi\bar\chi} v \rangle\,(\varepsilon^2 + 4 P)}\,,
 \end{equation}
and for sufficiently low temperature
i.e. for $x > \bar x_F$, Eq.(\ref{eq:delta}) becomes
\begin{equation} \label{eq:delta_late}
\frac{d \Delta_{\bar\chi}}{dx} = - \frac{\lambda \langle \sigma_{\chi\bar\chi} v
\rangle}{x\sqrt{x^2 + x^2_e}} \left( \Delta_{\bar\chi}^2 +
\varepsilon\Delta_{\bar\chi} \right)\,.
\end{equation}
We make the integration of Eq.(\ref{eq:delta_late}) from
$\bar{x}_F$ to $\infty$
and obtain the final WIMP abundance for anti--particles. The integration
range includes two parts: one is from
$\bar{x}_F$ to $x_e$ where the shear--dominated cosmology is used;
The second part is from $x_e$ to $\infty$ where the standard cosmology is
recovered. Then we obtain
\begin{equation} \label{eq:barY_cross}
Y_{\bar\chi}(x \rightarrow \infty) =  \frac{\varepsilon}
 { \exp \left[ 1.32\, \varepsilon \, m_{\chi} M_{\rm Pl}\,
 \sqrt{g_*} \,  I(\bar{x}_F, x_e)    \right] -1}\,,
\end{equation}
where
\begin{eqnarray}\label{eq:I}
  I(\bar{x}_F, x_e)   & =&
              \int^{x_e}_{\bar{x}_F} \frac{ \langle \sigma_{\chi\bar\chi} v \rangle }
              {x\sqrt{x^2 + x^2_e}}~ dx + \int^{\infty}_{x_e}
              \frac{ \langle \sigma_{\chi\bar\chi} v \rangle }
              {x^2}~ dx \nonumber \\ &=& \frac{a}{x_e}  {\rm ln}
              \left[\frac{x_e}{(1 + \sqrt{2})\bar{x}_F}
\left( 1 + \sqrt{1 + \left( \frac{\bar{x}_F}{x_e} \right)^2} \right) \right]
         \nonumber \\ & + & \frac{6b}{x^2_e}\,
   \left[- \sqrt{2} + \sqrt{1 + \left(\frac{x_e}{\bar{x}_F}\right)^2}\,\right]
         + \frac{a}{x_e} + \frac{3b}{x_e^2}\, .
\end{eqnarray}
Then the relic abundance for $\chi$ particle is
\begin{equation}\label{eq:Y_cross}
      Y_{\chi}(x \rightarrow \infty) = \frac{\varepsilon}
 {1 - \exp \left[- 1.32\, \varepsilon \, m_{\chi} M_{\rm Pl}\,
 \sqrt{g_*} \,  I(x_F, x_e)    \right] }\,,
\end{equation}
here $I(x_F, x_e)$ is same as Eq.(\ref{eq:I}) with the substitution of
$\bar{x}_F$ to
$x_F$. The two Eqs.(\ref{eq:barY_cross}) and (\ref{eq:Y_cross}) are
consistent with Eq.(\ref{eq:epsilon}) if $x_F = \bar x_F$.
In terms of the critical density $\rho_{\rm crit} = 3 M_{\rm Pl}^2 H_0^2$,
the final abundance is expressed as
\begin{equation}
\Omega_{\rm DM} h^2 =\frac{m_\chi s_0 \left[ Y_{\chi}~(x
  \rightarrow \infty) + Y_{\bar\chi}~(x \rightarrow \infty) \right] h^2}{\rho_{\rm  crit}}\,,
\end{equation}
where $s_0 = 2.9 \times 10^3~{\rm cm}^{-3}$ is the present
entropy density. The predicted present relic density for Dark Matter is then
given by
\begin{eqnarray} \label{omega}
 \Omega_{\rm DM}  h^2   =
               \frac{2.76 \times 10^8~ \varepsilon~m_\chi ~}
 {1 - \exp \left[- 1.32\, \varepsilon \, m_{\chi} M_{\rm Pl}\,
 \sqrt{g_*} \,  I(x_F, x_e)    \right] } -
 \frac{2.76 \times 10^8~ \varepsilon~m_\chi}
 {1 - \exp \left[ 1.32\, \varepsilon \, m_{\chi} M_{\rm Pl}\,
 \sqrt{g_*} \,  I(\bar{x}_F, x_e)    \right]  } \,.
\end{eqnarray}
The freeze--out temperature for $\bar{\chi}$ is fixed by assuming that the
deviation $\Delta_{\bar\chi} $ is of the same order of the equilibrium value of
$Y_{\bar\chi}$:
\begin{equation} \label{xf1}
\xi Y_{\bar\chi,{\rm eq}}( \bar{x}_F) = \Delta_{\bar\chi}( \bar{x}_F)\,,
\end{equation}
with $\xi$ being the numerical constant of order unity.
We choose the usual value $\xi = \sqrt{2} -1$ \cite{standard-cos}.
The obtained approximate analytic result matches with the exact numerical
result well within $5\%$ for all combinations of parameters.

%%%%%%%%%%%%%%%%%%%%%%%%%%%%%%%%%%%%%%%%%%%%%%%%%%%%%%%%%%%%%%%%%%%%%

\section{Constraints on parameter space}
\setcounter{footnote}{0}

We use the data in Eq.(\ref{data}) which is given by Planck
\cite{Ade:2015xua} to find the constraints on the parameter space
in shear--dominated universe for asymmetric Dark Matter.
\begin{figure}[h!]
  \begin{center}
    \hspace*{-0.5cm} \includegraphics*[width=8.7cm]{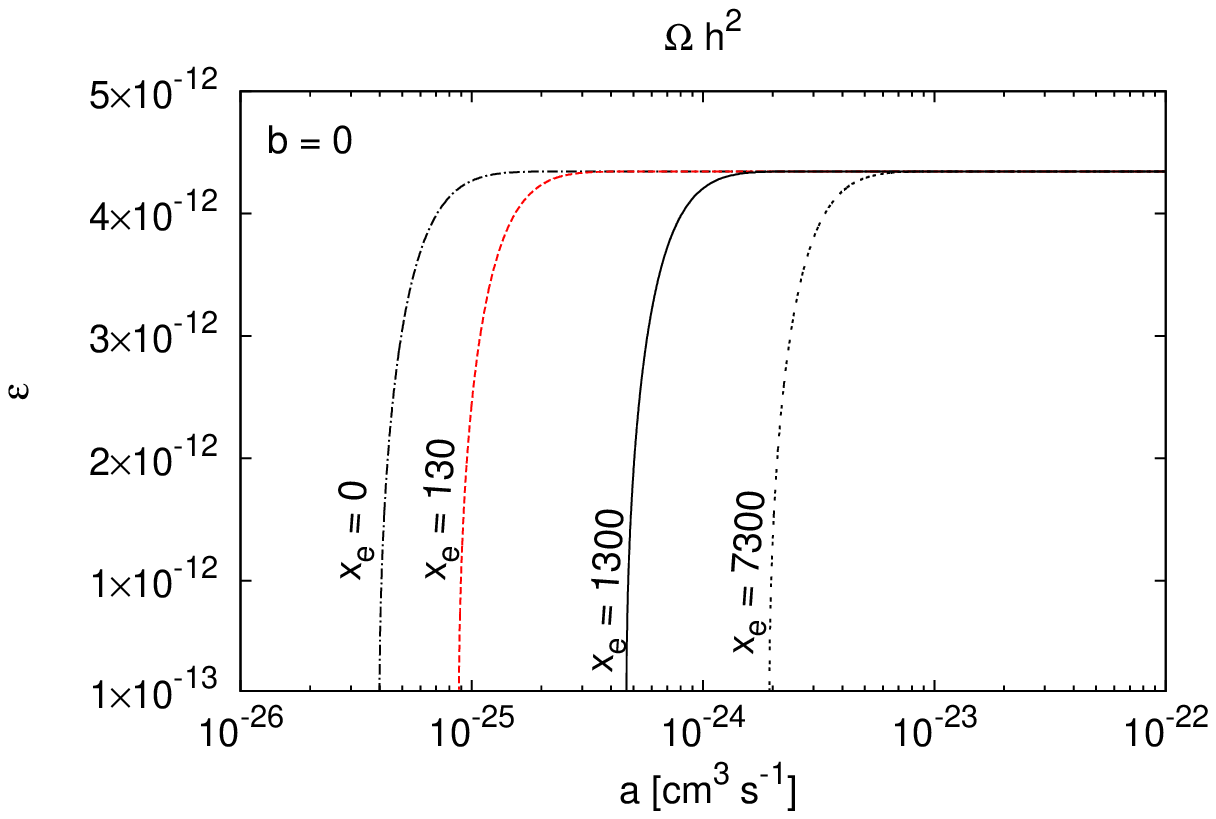}
    \put(-115,-12){(a)}
    \caption{\label{fig:d} \footnotesize
    The allowed region in the $(a,\varepsilon)$ plane for different $x_e$
    when $b=0$ and the Dark Matter relic density $\Omega_{\rm DM} h^2 = 0.1199$ .
    Here we take $m_\chi = 100$ GeV, $g_{\chi} = 2$ and $g_* = 90$.
    The dot--dashed line is for standard cosmology where $x_e = 0$ and the
    dashed (red) line
    is for $x_e = 130$, the thick line for $x_e = 1300$ and the double--dotted
    line for $x_e = 7300$.  }
    \end{center}
\end{figure}
Fig.\ref{fig:d} shows the relation between the cross section parameter
$a$ and asymmetry factor $\varepsilon$ for the shear--dominated universe
and standard cosmology when the Dark Matter relic density
$\Omega_{\rm DM} h^2 = 0.1199 $. The cross section for
shear--dominated model is larger than the standard cosmological model. This
is similar to the other non--standard cosmological models
\cite{Gelmini:2013awa,Iminniyaz:2013cla,Meehan:2014zsa,Wang:2015gua}. For
larger $x_e$, one needs larger annihilation cross
section. Indeed this can be understood from Fig.\ref{fig:a}. For
larger $x_e$, the particle and anti--particle decay more earlier than the
case for smaller $x_e$ and it results larger relic density. If one wants to
satisfy the observation range, the cross section must be large. In
Fig.\ref{fig:d}, i.e.
when $\varepsilon = 1\times 10^{-13}$, the s--wave
annihilation cross section $a = 3.99\times 10^{-26}$ ${\rm cm}^3 {\rm s}^{-1}$
for the standard cosmology which corresponds to the case $x_e =0 $ and it is
increased to $a = 8.82\times 10^{-26}$cm$^3$s$^{-1}$ for $x_e = 130$;
$a = 4.66\times 10^{-25}$cm$^3$s$^{-1}$ for $x_e = 1300$;
$a = 1.93\times 10^{-24}$cm$^3$s$^{-1}$ for $x_e = 7300$.

As we mentioned in the introduction, for asymmetric Dark Matter, in the
beginning, it is assumed there is less
anti--particle than the particle and the anti--particle is completely
annihilated away with the particle at late
time and there are only particles which made up the present total
Dark Matter
abundance in the standard cosmological scenario. Therefore, the asymmetric Dark
Matter particle is supposed to be detected by direct detection and the
indirect detection is not possible for the asymmetric Dark Matter.
However, it is different in nonstandard cosmological scenarios. The increase
of the annihilation cross section leads to the increase of the annihilation
rate. The increased
annihilation rate for particle and anti--particle gives the possibility to
detect the asymmetric Dark Matter in indirect way.
It is derived in Ref.\cite{Gelmini:2013awa}, the
ratio of the annihilation rate for asymmetric Dark Matter and symmetric Dark
Matter is less than 1,
\begin{equation}
      \frac{\Gamma_{\rm asym}}{\Gamma_{\rm Fermi}} =
 \frac{\langle \sigma_{\chi\bar\chi} v \rangle}{\langle \sigma v \rangle_{\rm Fermi}}
         \frac{2 Y_{\chi} Y_{\bar\chi}}{(Y_{\chi} + Y_{\bar\chi})^2} < 1 \,,
\end{equation}
Using Eq.(\ref{omega}), we obtain
\begin{equation}\label{limit}
       \frac{\Omega_{\rm DM} h^2}{2.76\times 10^8 m_{\chi}}
    \left(  1 - 2 \frac{\langle \sigma v \rangle_{\rm Fermi}}
     {\langle \sigma_{\chi\bar\chi} v \rangle} \right)^{1/2} < \varepsilon \,.
\end{equation}

\begin{figure}[h!]
  \begin{center}
    \hspace*{-0.5cm} \includegraphics*[width=8.7cm]{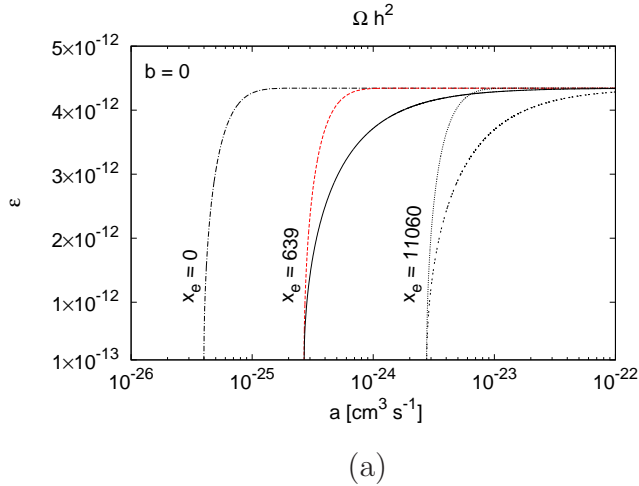}
    \put(-115,-12){(a)}
    \caption{\label{fig:e} \footnotesize
    The parameters are same as in Fig.\ref{fig:d} with the two lines
    without label are the limiting cross sections which corresponds to the
    Fermi--LAT bounds \cite{Ackermann:2013yva}
    where the thick line is for $\chi\bar\chi \rightarrow b\bar{b}$ and
    the double--dotted
    line is for $\chi\bar\chi \rightarrow \mu^{\dagger} \mu^{-}$; the dashed
    (red) line is for $x_e = 639$ and the dotted line is for
    $x_e = 11060$. The dot--dashed line is for standard cosmology.  }
    \end{center}
\end{figure}

Fermi Large Area Telescope (Fermi--LAT) gives
the upper bounds \cite{Ackermann:2013yva}
 $ a = 1.34 \times 10^{-25}$cm$^{3}$ s$^{-1}$ for
$\chi\bar\chi \rightarrow b\bar{b}$ and
$a = 1.38 \times 10^{-24}$cm$^{3}$ s$^{-1}$ for
$\chi\bar\chi \rightarrow \mu^{\dagger} \mu^{-}$ for
$m_{\chi} = 100$ GeV. Applying Planck data and Fermi-LAT data to Eq.(\ref{limit})
, we plot the limiting cross sections in shear--dominated model. The
thick line is for $\chi\bar\chi \rightarrow b\bar{b}$ and double--dotted line
is for $\chi\bar\chi \rightarrow \mu^{\dagger}\mu^{-}$ in Fig.\ref{fig:e}.
For the smaller asymmetry factor, i.e. $\varepsilon = 1 \times 10^{-13}$, the
corresponding limiting cross sections are
$\langle \sigma v \rangle_{\chi\bar\chi} \leq 2.68\times 10^{-25} $ for
$\chi\bar\chi \rightarrow b\bar{b}$ and
$\langle\sigma v\rangle_{\chi\bar\chi } = 2.76 \times 10^{-24}$ for
$\chi\bar\chi \rightarrow \mu^{\dagger}\mu^{-}$. For larger $\varepsilon$,
i.e. $\varepsilon = 3.5 \times 10^{-12}$, the corresponding limiting cross
sections are
$\langle \sigma v \rangle_{\chi\bar\chi} \leq 7.67\times 10^{-25} $ for
$\chi\bar\chi \rightarrow b\bar{b}$ and
$\langle\sigma v\rangle_{\chi\bar\chi } = 7.90 \times 10^{-24}$ for
$\chi\bar\chi \rightarrow \mu^{\dagger}\mu^{-}$.
The corresponding shear factor $x_e $ to the maximum annihilation cross
section must be $ 639$ for $\chi\bar\chi \rightarrow b\bar{b}$ and $ 11060$ for
$\chi\bar\chi \rightarrow \mu^{\dagger}\mu^{-}$ in shear--dominated universe.

However, for smaller value of $x_e$, when the cross section is large,
the anti--particle abundance is still suppressed as shown
in plot (d) in Fig.\ref{fig:a}. At finally, there is particle
only. Therefore, whether we can use the indirect detection signal for
asymmetric Dark Matter in shear--dominated universe depends on the value of
the shear factor $x_e$ and the annihilation cross section.

\section{Summary and Conclusions}

In this paper, we investigated the relic abundance of asymmetric
Dark Matter which decouples from the thermal equilibrium during the
nonstandard cosmological phase. We focus on the Bianchi type I model which
assumes the universe is expanded anisotropically before BBN. If the
freeze--out occurs
in the shear--dominated era, then the relic abundances of asymmetric Dark Matter
particle and anti--particle are affected. The Hubble rate is larger in
shear--dominated universe than the standard radiation--dominated cosmological
model. This leads to the early decay of the asymmetric Dark Matter particles.
As a result, the relic
abundances of Dark Matter particles increase. We found that the relic
densities of both particles and anti--particles are increased in
shear--dominated universe. The size of the increase depends on the value of
the shear factor $x_e$ and the annihilation cross sections. For the same cross
section, the increases are more sizable for larger $x_e$.

We used the Planck data to find the constraints on the annihilation cross
section and asymmetry factor $\varepsilon$ in shear--dominated model.
The cross section is increased for shear--dominated universe comparing to the
standard cosmology. For large $x_e$, the cross section should be large enough in
order to let the Dark Matter abundance falls in the observation
range. We showed that the
indirect detection signal is possible for asymmetric Dark Matter in
shear--dominated universe for appropriate $x_e$ value.
%The annihilation rate is large for asymmetric Dark Matter in
%nonstandard cosmological scenario.

If we know the annihilation cross section of the asymmetric Dark Matter
candidates and the annihilation rate is detectable at present, we can test the
universe before Big Bang Nucleosynthesis. Therefore, it is useful to
investigate the early universe before BBN.

\section*{Acknowledgments}

The work is supported by the National Natural Science Foundation of China
(11365022).

\end{document}